# Performance Stabilization of High-Coherence Superconducting Qubits

Andrew Dane, Karthik Balakrishnan, Brent Wacaser, Li-Wen Hung, H.J. Mamin, Daniel Rugar, Robert M. Shelby, Conal Murray, Kenneth Rodbell, and Jeffrey Sleight

IBM Quantum

Superconducting qubits have been used in the most advanced demonstrations of quantum information processing [1] [2] [3] [4], and they can be manufactured at-scale using proven semiconductor techniques [5] [6]. This makes them one of the leading technologies in the race to demonstrate useful quantum computers. Since their initial demonstration [7], advances in design [8] [9] [10] [11], fabrication [12], and materials [13] [14] [15] have extended the timescales over which fragile quantum information can be stored and manipulated on superconducting qubits. Ubiquitous atomic-scale material defects [16] [17] [18] [19] have been identified as a primary cause of qubit energy-loss and decoherence [20] [21]. Here we study transmon qubits that exhibit energy relaxation times exceeding 2.5 ms. Even at these long timescales, our qubit energy loss is dominated by two level systems (TLS). We observe large variations in these energy-loss times that would make it extremely difficult to accurately evaluate and compare qubit fabrication processes and to perform studies that require precise measurements of energy loss. To address this issue, we present a technique for characterizing qubit quality factor. In this method, we apply a slowly varying electric field to TLS near the qubit to stabilize the measured energy relaxation time, enabling us to replace hundreds of hours of measurements with ones that span several minutes.

Improving superconducting qubits requires that we accurately determine and compare the quality factor ($Q = 2\pi f_q T_1$) of devices formed using a variety of possible materials and processes. However, this task is challenging because of the time variation of two-level systems (TLS) loss [22]., It has been shown that accurately measuring the energy loss time ($T_1$) requires repeated measurements over long periods of time [23] [24]. Furthermore, it has been suggested that the variation in qubit $T_1$ increases as the mean $T_1$ increases [25]. To date, evaluating and comparing materials and fabrication processes to build better qubits necessitates the manufacture and time-consuming measurement of many qubits in an attempt to determine the average loss rate. Measuring qubit loss rates over time to estimate an average relies on sampling $T_1$ via the spectral diffusion of TLS with energies near the qubit frequency. This process is not fully understood but could be driven by thermal agitation of low energy TLS and TLS-TLS interactions [26] [27] [28], high energy particle bursts due to cosmic rays [29], or a variety of other athermal processes which could impact TLS with energies near the qubit energy [30]. Additionally, according to the standard tunneling model [17] [18], the frequencies of TLS can be modified by the application of strain [31] or electric field [32] [33].

Here, we use an electric field to modulate TLS frequencies during $T_1$ measurements to improve $T_1$ characterization of low-loss transmon qubits weakly coupled to readout resonators. The description of a $T_1$ measurement is provided in the Supplementary Information. By applying a voltage to a nearby TLS control electrode, we can modify the energies of TLS in the vicinity of the qubit, altering the qubit $T_1$. This can allow for a better estimation of the $T_1$ distribution of a given qubit.

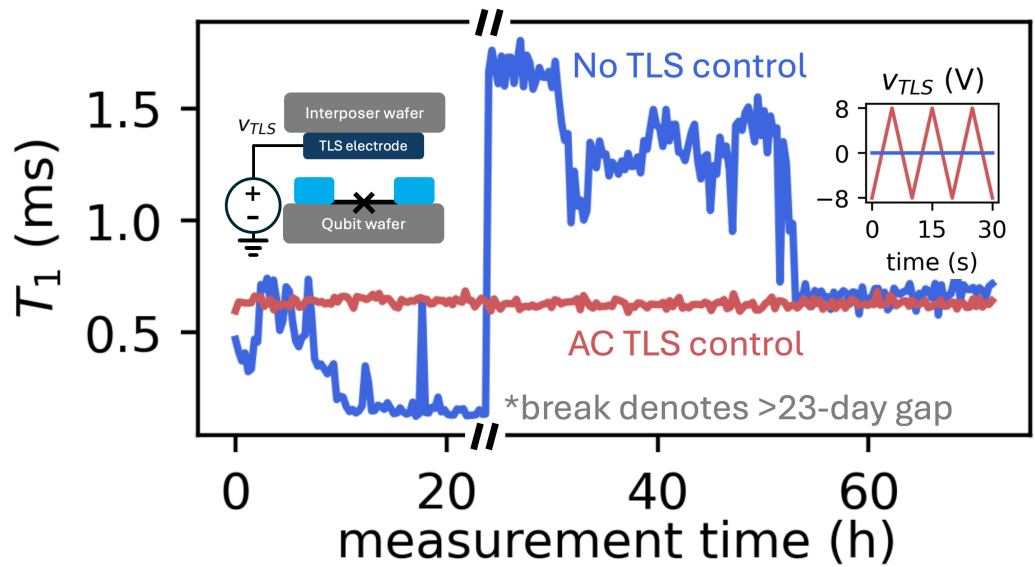

**Figure 1. Statement of the characterization problem and solutions provided by TLS control.** Interleaved measurements of $T_1$ over 26 days with TLS electrode set to 0V (blue) and with an AC waveform applied (red). Left inset: diagram of TLS control electrode. Right inset: applied TLS control waveforms.

The main results of our work rely on $T_1$ measurements made using three different types of voltage signals applied to a nearby TLS control electrode: (1) No-control, in which 0V was applied between the electrode and ground, (2) fast-random $T_1$ measurements, in which a randomly selected DC voltage between -8V and +8V was applied to the electrode for the duration of the $T_1$ measurement, and (3) AC-$T_1$ in which an asynchronous 0.1 Hz, 16 volt peak-to-peak triangle wave was applied to the electrode during the $T_1$ measurement. Fixed frequency transmons measured for this work possessed frequencies between 4.25 and 5.25 GHz and were dispersively coupled, with $\chi/2\pi \sim 200$ kHz, to readout resonators having $\kappa/2\pi \sim 100$ kHz and frequencies ranging from 6.5 to 7.5 GHz.

Figure 1 illustrates the difficulty in characterizing the $T_1$ of low-loss qubits through repeated measurements and previews our solution. We plot the results of interleaved no-control (dark blue) and AC (red) $T_1$ measurements. $T_1$ measurements were conducted over 24 hours, followed by a period of 23.3 days elapsing (during which time the sample remained cold) before an additional measurement period of about 48 hours. Each curve comprises approximately 240 $T_1$ values. $T_1$ fluctuations in the no-control time-series qualitatively illustrate why accurate $T_1$ characterization is difficult. For instance, if we were to use data extracted from a few hours of the no-control $T_1$ measurements to characterize the performance of this qubit, our conclusions would vary greatly depending on which period was selected. In contrast, AC $T_1$ measurements varied far less over the same period, offering a stable measurement in a much shorter time.

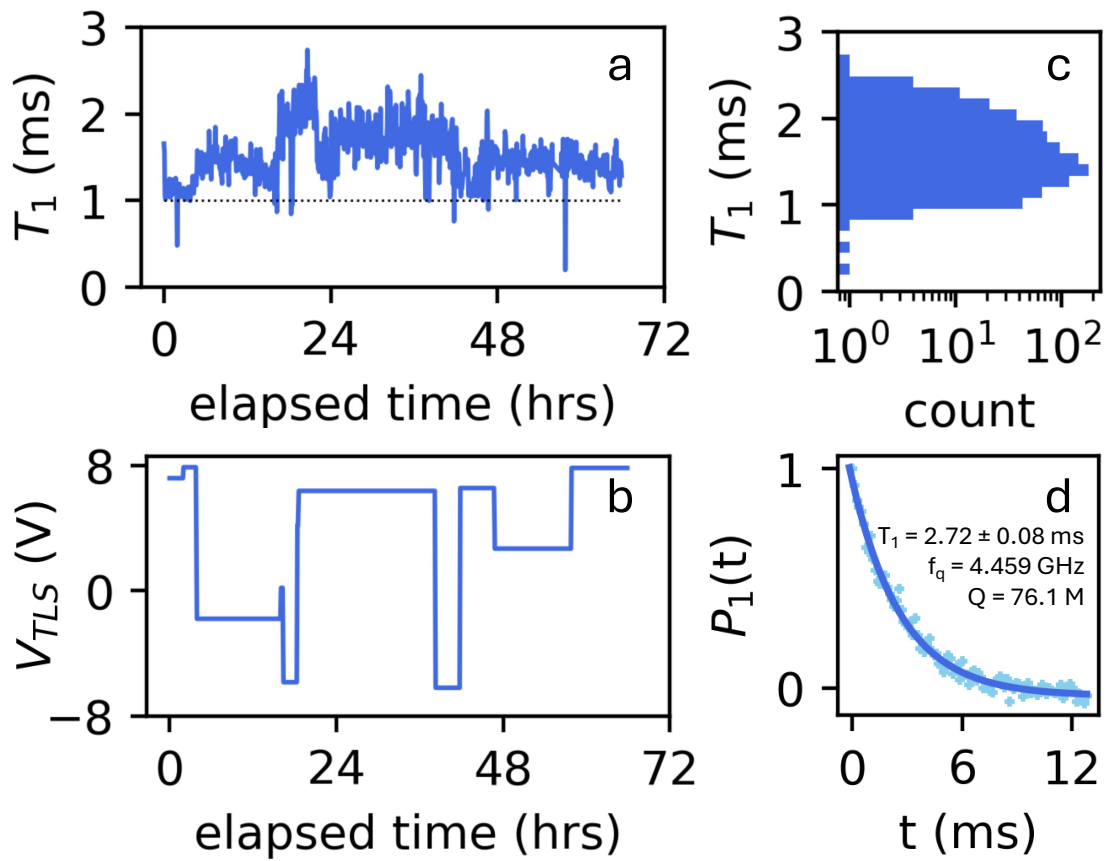

**Figure 2. Optimizing TLS control voltage to obtain long coherence times. (a)** Simple $T_1$ voltage-optimization protocol maintains $T_1$ longer than 1 ms (dotted line) for ~65 hours. **(b)** Applied TLS control voltage during the optimization period. **(c)** Resulting histogram of $T_1$ showing the vast majority of measurements exhibiting $T_1$ longer than 1 ms. **(d)** Measured $P_1(t)$ and exponential fit for the longest $T_1$ observed over all qubits.

We also demonstrate the implementation of TLS control on another qubit to optimize $T_1$ using a simple voltage optimization protocol (details in Supplementary Information). In Figure 2(a), we show that $T_1$ longer than 1 ms can be maintained using such an optimization protocol along with the utilized control voltages in Figure 2(b) and a histogram of obtained $T_1$ values in Figure 2(c). The longest $T_1$ measured using a similar protocol (details in Supplementary Information) is shown in Figure 2(d). This qubit had a frequency of 4.459 GHz, and based on this measurement, a quality factor, Q, of approximately 76 million was realized. This value provides a lower bound on the Q that would be possible if the effects of TLS under our control were mitigated, revealing an important piece of information that is typically not accessible without TLS control.

## TLS control and $T_1$ measurements

The main experimental results of this work consist of interleaved measurements of $T_1$ using the three types of TLS control previously mentioned, for 11 different qubits, over 72-93 hours of active data collection, spanning more than 26 days. However, first it is useful to briefly examine the results of a $T_1$ measurement in a way that will make it easier to interpret our findings.

$T_1$ (and the decay rate $\Gamma = 1/T_1$) varies continuously in time. In general, a time varying decay rate does not yield a $P_1(t)$ that is a simple exponential decay. To illustrate this point, consider the average of a family of exponential decay curves, whose decay rates follow a Gaussian distribution with mean μ and standard deviation σ. The resulting curve is known from probability theory [34]:

$$\langle e^{-\Gamma t} \rangle = \int e^{-\Gamma t} \cdot \Pr(\Gamma) \cdot d\Gamma = e^{-\mu t + \frac{1}{2}\sigma^2 t^2}$$

where $\Pr(\Gamma)$ represents the Gaussian probability density of the occurrence of decay rate $\Gamma$. When $t \ll \frac{2\mu}{\sigma^2}$, $\langle e^{-\Gamma t} \rangle \approx e^{-\mu t}$, arriving at a simple exponential with decay time $T_1 \approx \mu^{-1}$. Equivalently, $T_1$ can be described as the harmonic mean (h-mean) of the decay times whose inverses average to $\mu$. The distinction between arithmetic mean and harmonic mean is important when the distribution of $T_1$ contains a tail at low values. This distinction is not necessary in the AC-$T_1$ data, where mean and h-mean differ by less than one microsecond on average, but it is necessary for fast-random and no-control data, since they both usually contain low-fliers. Different distributions of $\Gamma$ yield different averaged curves, with no guarantee that those distributions can be approximated as a simple exponential. Nevertheless, these arguments motivate our comparison of the AC-$T_1$ value with the harmonic means of the fast-random and no-control data.

In Figure 3(a) we plot the results of interleaved AC (red), fast-random (light blue), and no-control (dark blue) $T_1$ measurements for one of eleven qubits measured in this manner (see the SI for the data sets of the other 10 qubits). At each step of the interleave, one AC, one no-control and four fast-random measurements were made. As depicted in Figure 1, after the first 24 hours of $T_1$ measurements, a break of 23.3 days occurred before measurements resumed. Each of the 11 data sets collected support what can be clearly seen in Figure 3(a): the AC $T_1$ measurement is far more stable in time than the either of other two. In Figure 3(b), we plot the cumulative harmonic means of the $T_1$ data in 3(a). The cumulative harmonic mean of the fast-random $T_1$ values (light blue) tends to converge to the AC harmonic mean more quickly than the no-control case. For all the measured qubits, the harmonic mean of the fast-random data was within 30% of the mean AC-$T_1$ data, and 7 of 11 were within 10%. Only 4 of 11 of the no-control data sets had a harmonic mean within 10% of the mean AC-$T_1$ at the conclusion of measurements. We take this finding as evidence that the $T_1$ that we measured in the fast-random experiments coincided with those that were sampled during an AC-$T_1$ measurement.

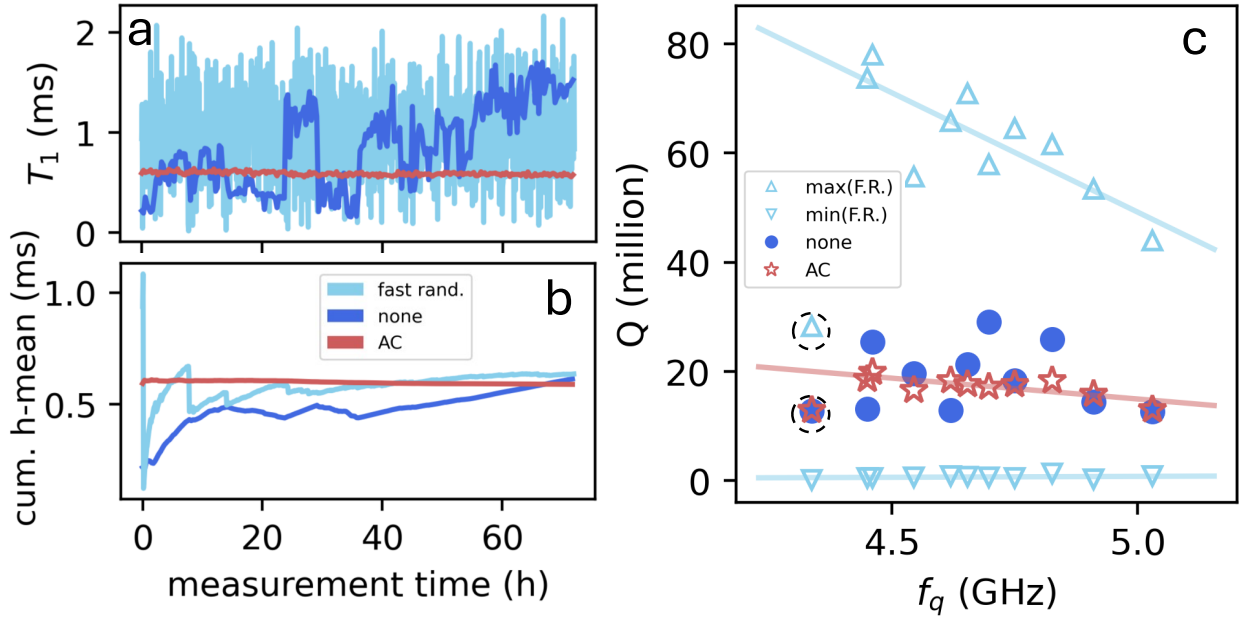

**Figure 3. $T_1$ measured with AC, fast-random, and no TLS control. (a)** Interleaved measurements of $T_1$ versus time and TLS control method: AC (red), fast-random (light blue), and no control (dark blue). Legend is shared with (b). Measurements were made in two periods, with a 23.3-day break after the initial 24 hours. 11 qubits were measured in a similar way with varying total measurement time. **(b)** Cumulative harmonic mean of the three data series in (a). The h-mean of the fast-random and no-control data appear to converge to the AC-$T_1$ value by the end of the measurement period. **(c)** Quality factor (Q) versus qubit frequency ($f_q$) for 11 qubits with Q calculated using: (1) the minimum and maximum fast-random $T_1$ (light blue triangles), (2) the h-mean of the $T_1$ measured without TLS control (dark blue circles), and (3) the h-mean of AC-$T_1$ measurements (red stars). The lines are simple linear fits (circled points not used).

Lastly, in Figure 3(c), we plot quality factor versus qubit frequency for each qubit. A decrease in Q with increasing frequency is evident in both the AC biased and maximum fast-random Q data. In the absence of TLS control methods, it is not clear if any relationship between Q and qubit frequency could be observed.

In Figure 4(a) and 4(b) we plot histograms of the $T_1$ data taken for the qubit with the shortest and longest mean AC-$T_1$, respectively. The measurements in Figure 4(a) exhibit both a lower mean AC-$T_1$, and a narrower spread in the measured fast-random data, as compared to the measurements depicted in Figure 4(b). In Figure 4(c) we plot the standard deviation of the full time series of $T_1$ for each TLS control scheme, as a function of the mean of the AC-$T_1$ measurement of the same qubit, for all 11 qubits. All three distributions show a positive correlation between the mean of the AC measurements, and the standard deviation of $T_1$ measured with the given TLS control, demonstrating that the variation in qubit $T_1$ increases as $T_1$ increases [25]. In all cases the standard deviation of the AC-$T_1$ measurements is far lower than the other two.

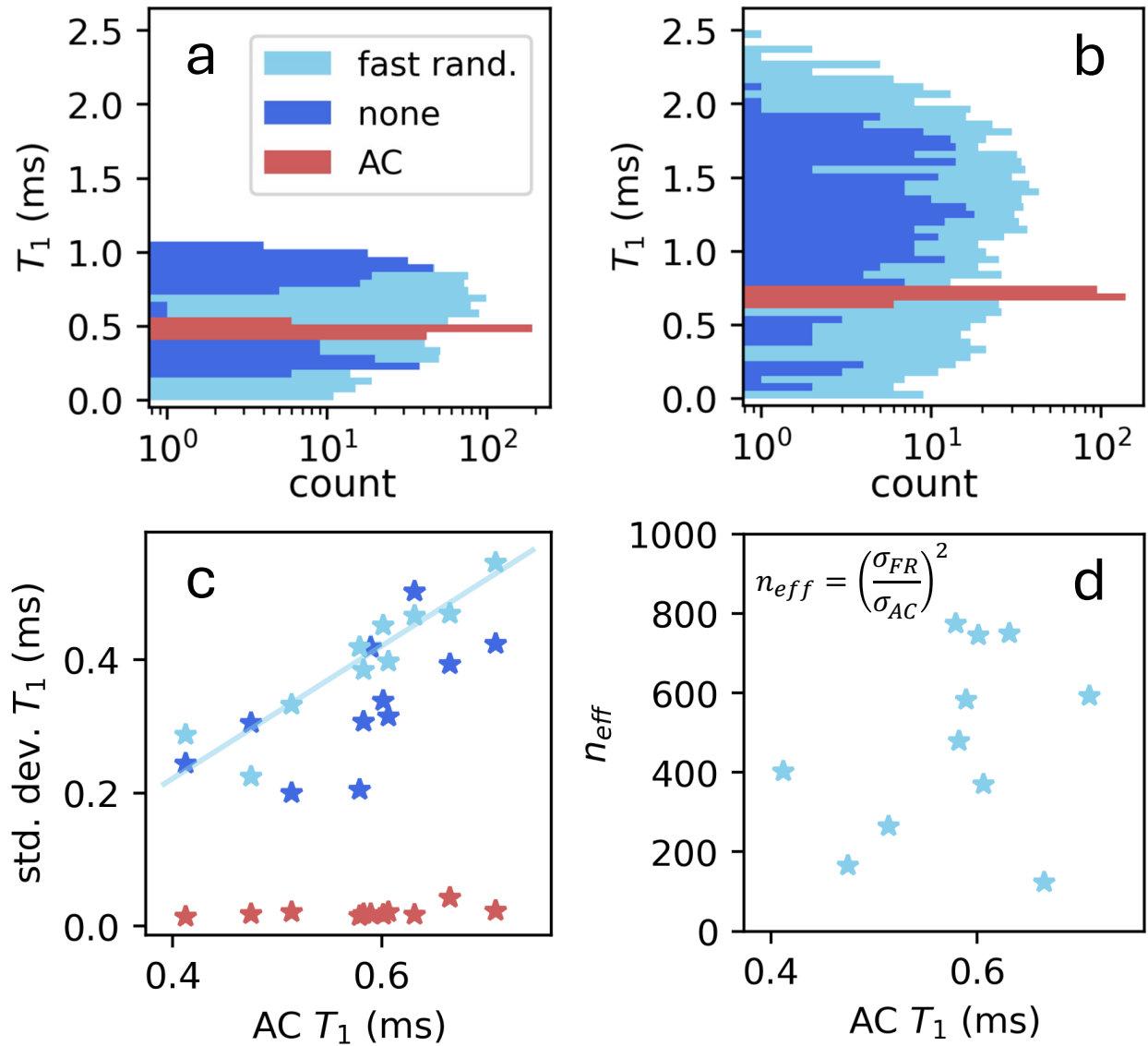

**Figure 4. The distribution of $T_1$, revealed by and controlled with applied voltage (a-b)** Histograms of $T_1$ measured using each TLS control scheme, for the qubit with the shortest and longest average $T_1$ among the 11 qubits measured as in Figure 3(a). To determine the shortest average $T_1$, we did not consider the qubit which was off trend in 3(c). The legend in (a) applies to (a-c). **(c)** The standard deviation of the measured $T_1$ for each type of TLS control. The fitted line has a slope of $\sigma/\mu$=1.0. **(d)** Number of identical qubits, $n_{eff}$, required to reach a commensurate reduction in the standard deviation of $T_1$ as enabled by AC control for each of the 11 measured qubits.

If we assume that the AC-$T_1$ and fast-random $T_1$ measurements are unbiased estimators of the $T_1$ mean and standard deviation, respectively, we can rewrite the standard error of the AC-$T_1$ ($\sigma_{AC}$) as $\sigma_{AC} = \frac{\sigma_{FR}}{\sqrt{n}}$, where $\sigma_{FR}$ is the sample standard deviation of the fast-random $T_1$ distribution. After rearranging to $n = \left(\frac{\sigma_{FR}}{\sigma_{AC}}\right)^2$, for illustrative purposes, if we assume that each qubit has an independent and identical $T_1$ distribution, then *n* can represent a number of qubits rather than a number of measurements. Thus, we take *n* to be the "effective number of qubits" ($n_{eff}$) required to reach a commensurate reduction in the standard deviation of $T_1$ as enabled by AC control. Figure 4(d) is a plot of $n_{eff}$ for each of the 11 qubits showing that, on average, hundreds of qubits would be needed to obtain standard error of a single AC-$T_1$ measurement.

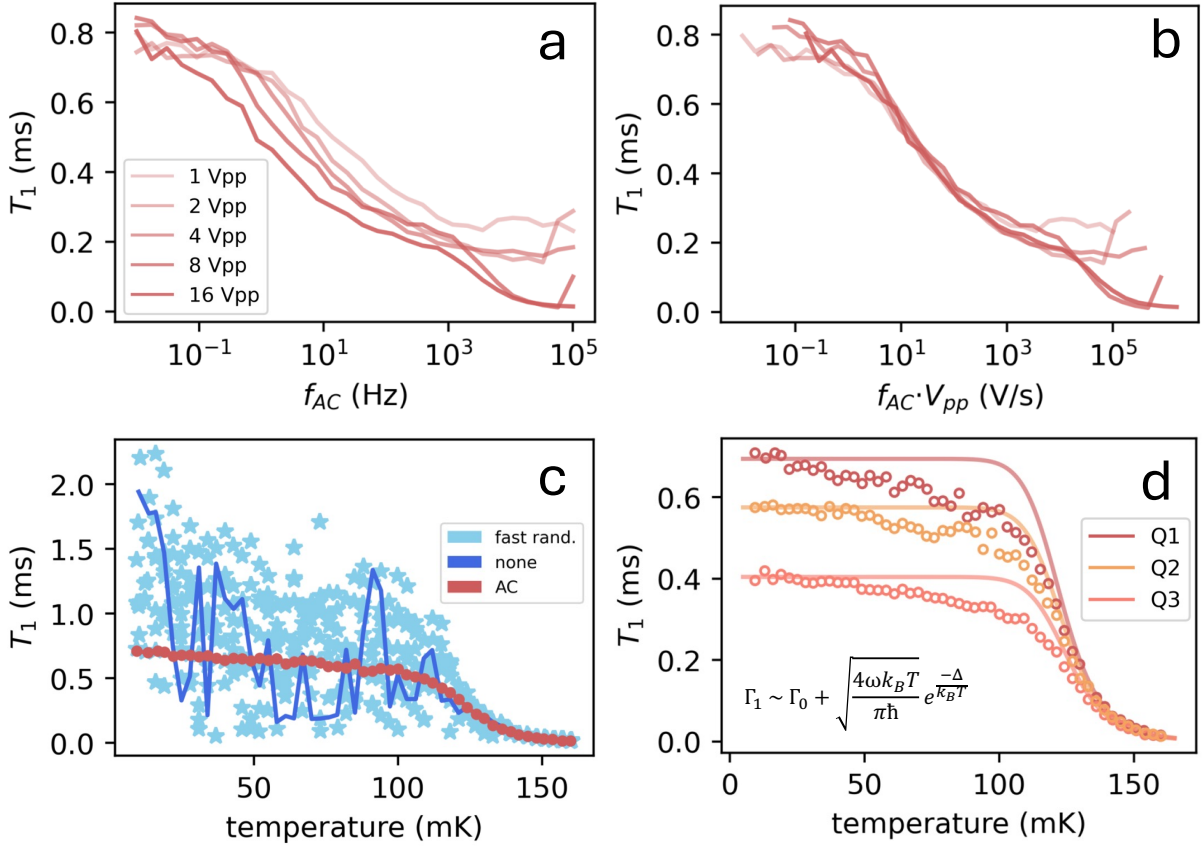

**Figure 5. Using TLS control to improve our understanding about qubit loss times (a)** AC-$T_1$ measured using different peak-to-peak voltages and AC waveform frequencies ($f_{AC}$). Legend shared with (b). **(b)** AC-$T_1$ from (a) plotted vs. $f_{AC} \cdot V_{pp}$. **c)** $T_1$ measured using AC, fast-random and no-control of the TLS as a function of mixing chamber temperature. **(d)** AC-$T_1$ of three different qubits versus temperature, revealing a linear dependence of $T_1$ at low temperatures and significant degradation above 100 mK due to thermal quasiparticles. Inset: equation used to fit AC-$T_1$ vs. temperature data.

In Figure 5 we highlight two additional characterizations enabled by AC TLS control. In Figure 5(a) we plot AC-$T_1$ measured using different peak-to-peak voltages ($V_{pp}$) and AC frequencies ($f_{AC}$) and find that in all cases the $T_1$ is reduced as the frequency is increased. When we plot the same $T_1$ data as a function of Volts/second due to the combination of $V_{pp}$ and frequency, these curves overlap. Similar behavior was observed in high-Q superconducting resonators connected to a network of TLS-containing capacitors that could be voltage biased at low frequencies [35]. Those measurements were explained by Landau-Zener transitions in the combined resonator-TLS system, with increased likelihood as the TLS energies were swept faster at higher voltage sweep rates [36].

Further, we demonstrate how AC TLS control can be used to precisely measure qubit energy loss and reveal subtle trends. In Figure 5(c) we plot the results of running the no-control, fast-random, and AC control measurements versus temperature on one qubit. The temperature was PID controlled by applying power to a resistive heater on the sample stage of our dilution refrigerator. The AC-$T_1$ data is far less noisy than the other data sets until loss from thermal quasiparticles become prominent around 100 mK. In Figure 5(d) we plot AC-$T_1$ versus temperature for three different qubits along with fit lines. The fitting assumes that $\Gamma_1$ ($= T_1^{-1}$) is the sum of a constant loss rate ($\Gamma_0$) plus a temperature dependent term due to thermally generated quasiparticles [37] [38]. For a given fit, the inverse of $\Gamma_0$ is set equal to the average of the AC-$T_1$ measurements below ~40 mK, and only the gap (Δ) was used as a fitting parameter, with points T~<135 mK excluded from the fit to ensure good agreement at higher temperature. The observed discrepancy between the fits and data at intermediate temperatures illustrates an additional component of qubit relaxation that would be difficult to distinguish without the AC measurements.

# Conclusions

We presented a new method for characterizing superconducting qubit quality factor and demonstrated it using qubits that exhibited $T_1$ times exceeding 2.5 ms. By applying a low frequency AC voltage to an electrode near the qubit, we significantly decreased the variation in measured $T_1$ values. We conclude that AC TLS control is a powerful technique to enable rapid and accurate characterization of qubits to allow for comparisons among (for example) different fabrication processes. Furthermore, we showed that as qubit $T_1$ improves, the variations in $T_1$ also increase, underscoring the increasing utility of AC control as relaxation times improve into the millisecond regime. Finally, we used AC TLS control to conduct precise measurements of energy loss as a function of qubit frequency and temperature. These measurements revealed subtle trends that were difficult to detect without TLS control. Going forward, we hope this new method of AC TLS control will be adopted by the broader technical community to both accelerate improvements in qubit quality factor and obtain new insights into the physics of loss mechanisms that impact superconducting qubits.

# Acknowledgements

The authors thank Joseph Finley, Jim Hannon, Timothy Phung, Martin Sandberg, and Jerry Tersoff for prior insights covering both simulations and measurements, and help with many aspects of these evolving experiments. The authors also thank Kenny Tran for various technical assistance. Samples were fabricated using the IBM Microelectronics Laboratory and the IBM Central Scientific Services facility.

# Supplementary Information

## $T_1$ Measurement Description

A single $T_1$ measurement consisted of measuring the probability of the qubit remaining in the excited state ($P_1$), conducted at a series of fixed delay times after initialization in the |1> state, and then fitting the resulting decay curve to a falling exponential. Each $P_1$ was the compilation of at least 400 identical excitation-delay-measure rounds. The time delays were logarithmically spaced to resolve decay curves having a wide range of $T_1$ times with a fixed set of delays. No-control and AC $T_1$ measurements used 81-101 delay times, while the fast random used 21-25. Each No-control and AC $T_1$ measurement took ~10 min., and each fast random $T_1$ measurement about 2.5 min.

## $T_1$ Optimization Protocol

A $T_1$ measurement was performed with a randomly chosen voltage held on the TLS electrode. Each time the resulting $T_1$ was longer than 1 ms, the same voltage was used for the next $T_1$ measurement. Each time the resulting $T_1$ was shorter than 1 ms, a new voltage was chosen at random and a $T_1$ measurement was performed. This sequence was repeated until the measured $T_1$ was longer than 1 ms, after which the new voltage was used for subsequent $T_1$ measurements. For the results shown in Figure 2(a-c), all instances in which a new random voltage was necessary to meet the threshold criteria of $T_1 > 1$ ms required 3 or fewer attempts.

To obtain the plot shown in Figure 2(d), first, a fast-random $T_1$ measurement was performed. If the resulting $T_1$ was greater than 2 ms, then the randomly selected voltage was held on the TLS electrode. Then, a finer $T_1$ measurement was made, with 100 linearly spaced time delays and a maximum time delay equal to 4 times the fast random $T_1$ at the same voltage. Afterward the process was repeated with a new fast-random measurement at another randomly selected voltage.

# Supplementary Figures

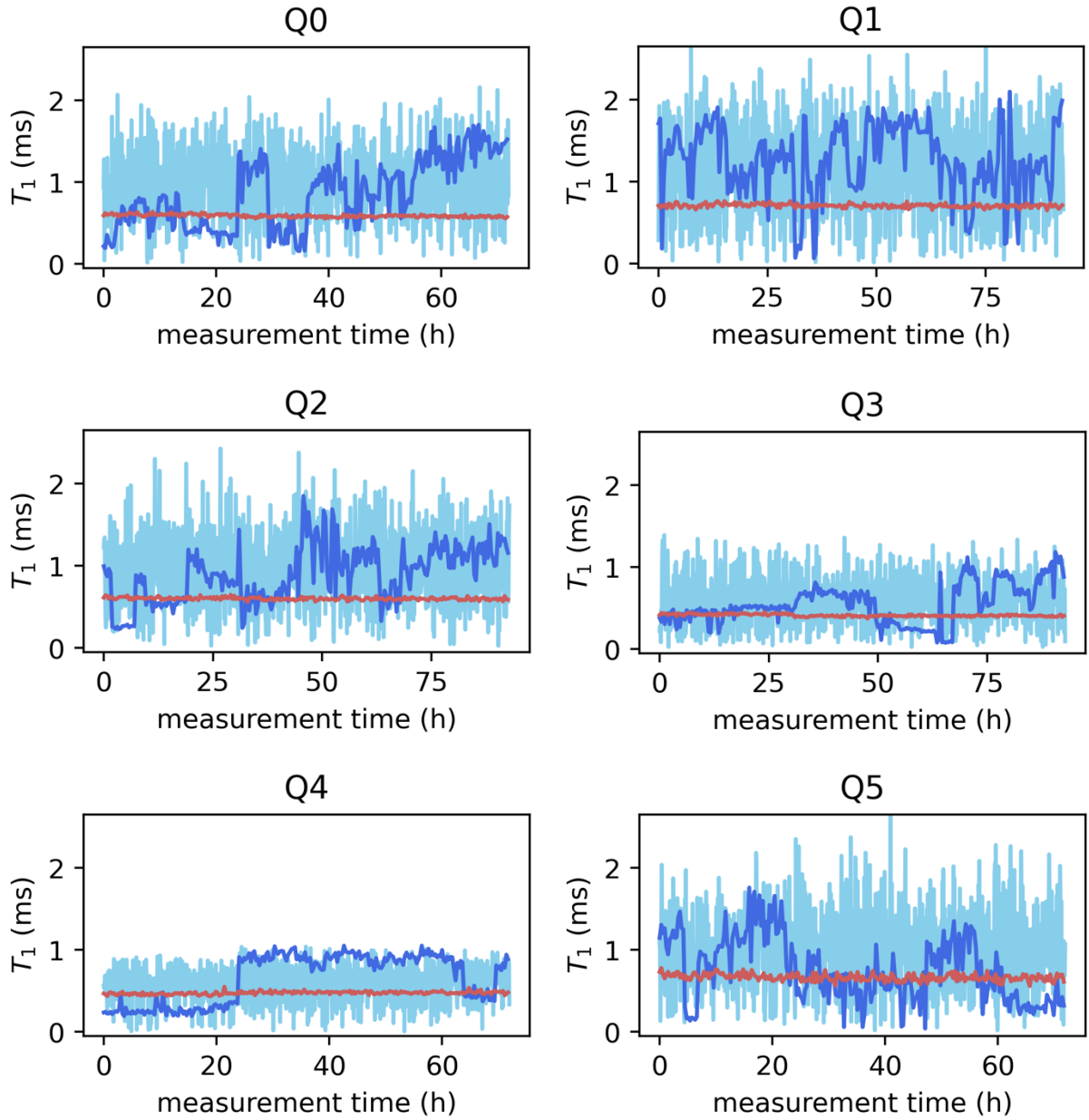

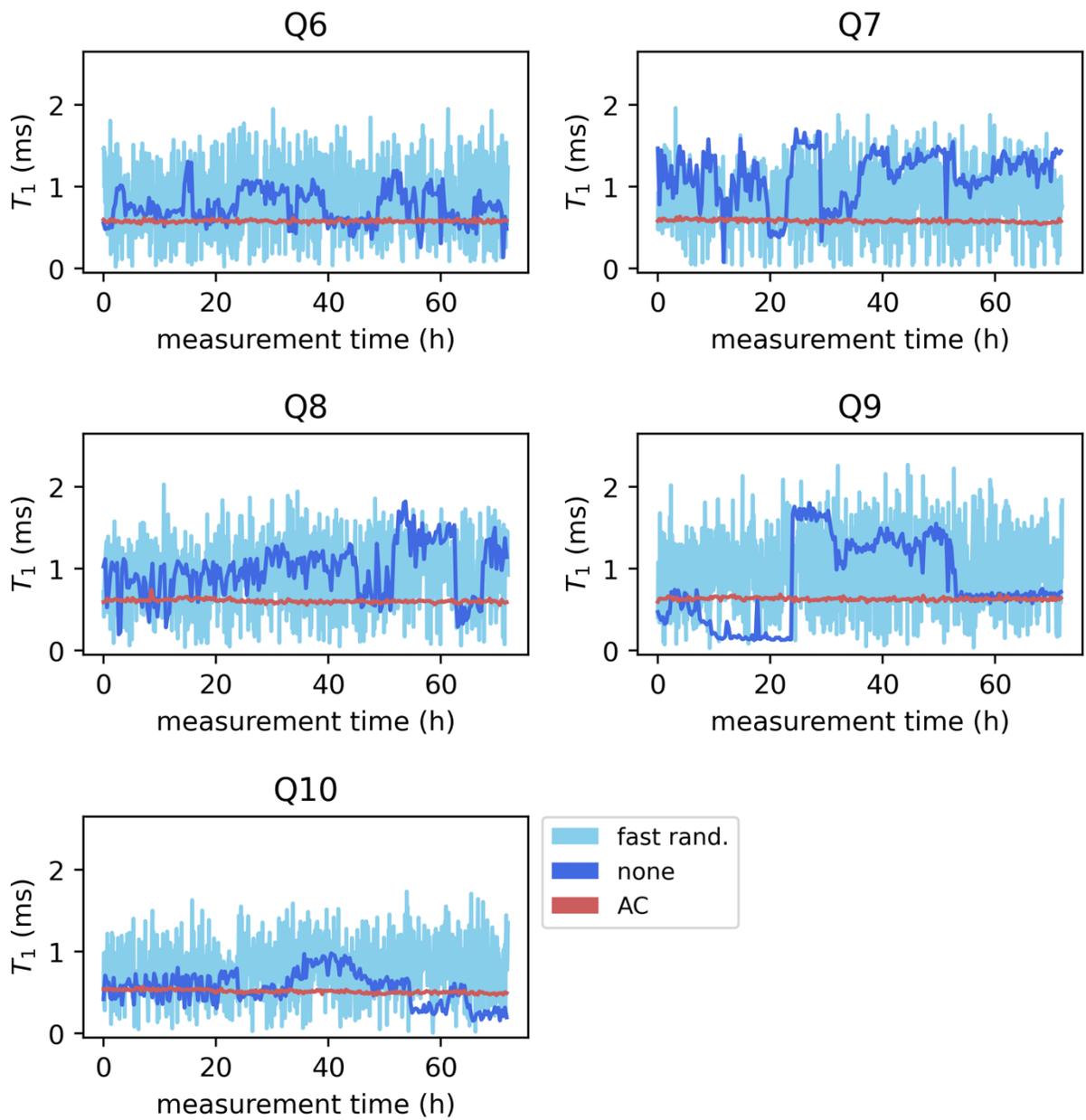

**Figure S1. Full dataset of $T_1$ measured over time with AC, fast-random and no-control.**